\documentclass[11pt]{article}
\usepackage{amsmath}
\usepackage{graphicx}
\usepackage{amscd}
\newtheorem{thm}{Theorem}

 \newtheorem{lem}{Lemma}
 \newtheorem{prop}{Proposition}
\newtheorem{rem}{Remark}
\begin{document}
\title{Darboux transformation for two component derivative nonlinear Schr\"odinger
equation}
\author{Liming Ling and Q.~P.~Liu\\[5pt]
Department of Mathematics\\
China University of Mining and Technology\\
Beijing 100083, P R China}
\date{}
\maketitle


\begin{abstract}

In this paper, we consider the two component derivative nonlinear Schr\"{o}dinger equation and present a simple
Darboux transformation for it. By iterating this Darboux transformation, we construct a compact representation  for the $N-$soliton solutions.
\end{abstract}

\maketitle

\section{Introduction}
The nonlinear partial differential  equations with multi-soliton solutions have been studied extensively. They  are often
widely applicable in physics and thus constitute very important equations in mathematical physics. The celebrated examples include Korteweg-de Vries equation, sine-Gordon equation and nonlinear Schr\"{o}dinger (NLS) equation and many others. These systems, named as soliton or integrable equations, are also very rich in mathematical properties and whole subject is closely related to other mathematical branches such as differential geometry, algebraic geometry, combinatorics, Lie algebras, etc \cite{dickey}.

Since integrable systems have remarkable mathematical properties and numerous physical applications, their generalizations or extensions have attracted  attention of many researchers. One possible direction is multi-component generalization. This sort of extensions may also be physically interested. The most famous example might be the Manakov's two component NLS equation, which now is one of the most  important equations in theory of pulse propagation along the optical fiber.

Another interesting soliton equation is the derivative nonlinear
Schr\"{o}dinger (DNLS) equation
\[
iq_t=-q_{xx}+\frac{2}{3}i\epsilon (|q|^2q)_x,
\]
which appeared in Plasma physics (see \cite{Rog}\cite{mj}), describing
Alfv\'{e}n wave propagation along the magnetic field. This equation
was solved by inverse scattering transformation by Kaup and Newell
\cite{kaup}. Much  research has been conducted for it and many
results have been achieved. We mention here a simple looked Darboux
transform, obtained independently by Imai \cite{inami} and Steudel \cite{ste}, enables one
to get its explicit $N-$soliton solution. The two component
extension of DNLS equation was constructed by Morris and Dodd
\cite{md}. It reads as
\begin{eqnarray}
\label{eq:y1} iq_{1t}&=&
-q_{1xx}+\frac{2}{3}i\epsilon\left[(|q_1|^2+|q_2|^2)q_1\right]_x,
\\[1pt]
\label{eq:y2} iq_{2t}&=&
-q_{2xx}+\frac{2}{3}i\epsilon\left[(|q_1|^2+|q_2|^2)q_2\right]_x,
\end{eqnarray}
where $\epsilon=\pm 1$. This system was studied by means of inverse
scattering transformation. For convenience, we take $\epsilon=-1$ in
the subsequent discussion.  The zero-curvature representation in
this case reads as
\begin{eqnarray}\label{eq:zero1}
   \Phi_x
   &=& U\Phi,\\[2pt]\label{eq:zero2}
\Phi_t
   &=& V  \Phi,
\end{eqnarray}
where $\Phi=\begin{pmatrix}
     \phi_1,
     \phi_2,
     \phi_2
   \end{pmatrix}^T$, $\zeta$ is the spectral parameter and
\[
U=U_2\zeta^2+U_1\zeta,\;
V=\zeta^4 V_4+\zeta^3 V_3+\zeta^2
V_2+\zeta V_1
\]
with
\[
U_2=\begin{pmatrix}
             -2i & 0 & 0 \\
             0 & i & 0 \\
             0 & 0 & i \\
           \end{pmatrix},\;\;
U_1=\begin{pmatrix}
          0 & q_1 & q_2 \\
          r_1 & 0 & 0 \\
          r_2 & 0 & 0 \\
        \end{pmatrix},\;\;
V_4=\begin{pmatrix}
          -9i & 0 & 0 \\
          0 & 0 & 0 \\
          0 & 0 & 0 \\
        \end{pmatrix},
        \]
        \[
V_3=\begin{pmatrix}
      0 & 3q_1 & 3q_2 \\
      3r_1 & 0 & 0 \\
      3r_2 & 0 & 0 \\
    \end{pmatrix},\;\;
V_2=\begin{pmatrix}
      -i(r_1q_1+r_2q_2) & 0 & 0 \\
      0 & ir_1q_1 & ir_1q_2 \\
      0 & ir_2q_1 & ir_2q_2 \\
    \end{pmatrix},
    \]
    \[
V_1=\begin{pmatrix}
         0 & iq_{1x}+\frac{2}{3}(r_1q_1+r_2q_2)q_1 &iq_{2x}+\frac{2}{3}(r_1q_1+r_2q_2)q_2 \\
          -ir_{1x}+\frac{2}{3}(r_1q_1+r_2q_2)r_1 & 0 & 0 \\
          -ir_{2x}+\frac{2}{3}(r_1q_1+r_2q_2)r_2) & 0&
          0 \\
        \end{pmatrix}.
\]
Then a straightforward calculation shows that the compatibility
condition of  \eqref{eq:zero1}-\eqref{eq:zero2} leads to a system
which reduces to \eqref{eq:y1} and \eqref{eq:y2} under condition
$r_k^*=-q_k$ $(k=1,2)$.

The purpose of this paper is to construct a compact representation
of the $N-$soliton solution for the  two component DNLS equation.
We shall take Darboux transformation approach. Indeed, the original Darboux transformation, which is associated with
Sturm-Louiville equation, has been generalized to many other differential and difference equations. It turns out that this approach
often leads to nice representations in terms of determinants for solutions of nonlinear systems
and thus an ideal method to construct $N-$soliton solutions (see \cite{mat}\cite{gu}\cite{lable}\cite{ces}). In particular, Darboux transformations for certain multi-component integrable equations have been studied in \cite{pa:O}\cite{pa:Q}.

The paper is organized as follows. In next section, we construct an elementary Darboux transformation for the general system \eqref{eq:zero1}-\eqref{eq:zero2},
which naturally induces a Darboux transformation for the conjugate system. Then, we combine two Darboux transformations together and find a two-fold Darboux transformation, which turns out to be the proper one for the reduction we are interested in. The reduction problem will be tackled in the section 3 and an elegant Darboux transformation will be given there for our two component DNLS equation. In section 4, we iterate our Darboux transformation and give N-soliton solutions of  two component DNLS equation in terms of determinants. Final section includes some discussion.

\section{Darboux transformation in general}
We now consider the general linear system \eqref{eq:zero1}-\eqref{eq:zero2} and manage to find a Darboux transformation for it. Our strategy is to find a proper Darboux transformation such that it can be easily reduced to the two-component DNLS case.  To this aim, we start with an elementary
Darboux transformation
\[
\hat{\Phi}=T_1\Phi
\]
with Darboux matrix $T_1=\zeta
T_{11}+T_{10}$. After some calculations and analysis, we find that $T_1$ has to take the following explicit form
\begin{eqnarray}\label{eq:T0}
T_1 &=&\begin{pmatrix}
               a\zeta & c_1 & c_2 \\
               c_3 & b\zeta & c\zeta \\
               c_4 & d\zeta & e\zeta \\
             \end{pmatrix},
\end{eqnarray}
where ~$a, b, c, d$ and  $e$ are the functions of $(x,t)$, while $c_1,c_2,c_3$ and $c_4$ need to be
constants. For convenience, we make the assumption
\begin{equation}\label{c1}
c_1=c_3=1,c_2=c_4=0.
\end{equation}

Since $\mathrm{Tr}(U)=0$, $\mathrm{Tr}(V)=-9i\zeta^4$, we may assume
\begin{equation}\label{eq:2}
   \det(T_1)=\zeta(\frac{\zeta^2}{\zeta_1^2}-1)
\end{equation}
where ~$\zeta_1$ is a complex constant. Thus, the Darboux matrix $T_1$ is singular at $\zeta=\zeta_1$. Next we associate the entries of $T_1$ with a special solution of our linear systems \eqref{eq:zero1}-\eqref{eq:zero2}. To this end, taking ~$ (\varphi_1, \varphi_2, \varphi_3)^{T}$ as a corresponding solution of the Lax pairs at ~$\zeta=\zeta_1$ and requiring
\begin{equation}\label{eq:1} T_1|_{\zeta=\zeta_1}\begin{pmatrix}
                   \varphi_1\\
                   \varphi_2\\
                   \varphi_3\\
                 \end{pmatrix}=0,
\end{equation}
we obtain
\begin{equation}\label{a}
 a = -\frac{\varphi_2}{\zeta_1\varphi_1}, \;
 b = \frac{-\varphi_1+\zeta_1Q_2\varphi_3}{\zeta_1\varphi_2},\;
 c= -Q_2, \;
 d = -\frac{\varphi_3}{\varphi_2},\; e=1.
\end{equation}
where $Q_2$ is a potential: $ Q_{2,x}=q_2$.

Now we have the following
\begin{thm}
Let $ (\varphi_1, \varphi_2, \varphi_3)^{T}$ be a particular solution of \eqref{eq:zero1}-\eqref{eq:zero2} at $\zeta=\zeta_1$ and
the matrix $T_1$ be given by (\ref{eq:T0}) with entries defined by \eqref{c1} and \eqref{a}. Then $T_1$ is a Darboux matrix for  the linear system
\eqref{eq:zero1})-(\eqref{eq:zero2}, namely $\hat{\Phi}=T_1\Phi$ is a new solution of  \eqref{eq:zero1})-(\eqref{eq:zero2}). The transformations between fields are given by
\begin{align*}
 & \hat{q}_1 =  \frac{(q_1\varphi_2+q_2\varphi_3-3i\zeta_1\varphi_1)\varphi_2}{\varphi_1^2},&
  \hat{r}_1 =  \frac{(r_1\varphi_1+3i\zeta_1\varphi_2)\varphi_1+(r_2\varphi_2-r_1\varphi_3)\zeta_1Q_2\varphi_1}{\varphi_2^2},\\[8pt]
 &  \hat{r}_2 =  \frac{(r_1\varphi_3-r_2\varphi_2)\zeta_1\varphi_1}{\varphi_2^2}, &\hat{q}_2 =  \frac{(\varphi_2q_1+\varphi_3q_2-3i\zeta_1\varphi_1)\zeta_1Q_2\varphi_2-q_2\varphi_1\varphi_2}{\zeta_1\varphi_1^2},
\end{align*}
where hatted quantities are transformed variables.
\end{thm}

\noindent
\textbf{Proof:} What we need to do is to check that the following equations
\begin{equation*}
    T_{1x}+T_1U=\hat{U}T_1,\qquad T_{1t}+T_1V=\hat{V}T_1
\end{equation*}
hold. Where
\[
\hat{U}=\zeta^2U_2+\zeta \hat{U}_1,
\hat{V}=\zeta^4V_4+\zeta^3\hat{V}_3+\zeta^2\hat{V}_2+\zeta\hat{V}_1
\]
and  $\hat{U}_1$, $\hat{V}_3$, $\hat{V}_2$ and  $\hat{V}_1$ are
$U_1$, $V_3$, $V_2$, $V_1$ with the corresponding entries $r_1$, $r_2$, $q_1$ and  $q_2$
are replaced respectively by $\hat{r}_1$, $\hat{r}_2$,$\hat{q}_1$ and
$\hat{q}_2$. Checking can be done by direct calculations.


\begin{rem}\label{rm:1}
It is interesting to note that under this Darboux transformation, we also have an alternative representation for $\hat{r}_2$:
 $d_x=\hat{r}_2$.
\end{rem}

To proceed, we notice that the two component DNLS equation also has the following Lax pairs
\begin{eqnarray}\label{con1}
-\Psi_x
   &=&\Psi U,\\ \label{con2}
-\Psi_t
   &=& \Psi V,
\end{eqnarray}
where $\Psi=\begin{pmatrix}
     \phi_1 ,& \phi_2 ,& \phi_3
   \end{pmatrix}$ and $U$ and $V$ are as above. This linear problem actually is the conjugate problem of
\eqref{eq:zero1}-\eqref{eq:zero2}. A simple but useful observation is
\begin{lem}\label{tm:2}
If the matrix $T$ is a Darboux matrix of the original linear system \eqref{eq:zero1}-\eqref{eq:zero2}, then $T^{-1}$ is a Darboux matrix of the conjugate
linear system \eqref{con1}-\eqref{con2}.
\end{lem}
\noindent
\textbf{Proof}: Direct calculation.

\bigskip

Now we consider the conjugate linear system and its Darboux transformation. The analysis goes as in the case of the original linear system: Taking
 $(\chi_1,\chi_2,\chi_3)$ as a special solution of the system \eqref{con1}-\eqref{con2} at $\zeta=\zeta_2$, and constructing the following matrix
\begin{equation}\label{Tt}
T_2=\begin{pmatrix}
               \hat{a}\zeta & 1 & 0\\
               1& \hat{b}\zeta & \hat{c}\zeta \\
               0& \hat{d}\zeta &  \zeta \\
    \end{pmatrix},
\end{equation}
where
\begin{eqnarray} \label{eq:2b}
  \hat{a} = -\frac{{\chi}_2}{\zeta_2{\chi}_1},\;
  \hat{b} = -\frac{{\chi}_1+\zeta_2R_2{\chi}_3}{\zeta_2{\chi}_2},\;
  \hat{c} = -\frac{{\chi}_3}{{\chi}_2},\; \hat{d}=R_2
\end{eqnarray}
and
\[
R_{2,x}=r_2,
\]
we have
\begin{thm}
The matrix $T_2$ defined by \eqref{Tt} is an elementary Darboux matrix of the
conjugate linear system \eqref{con1}-\eqref{con2} and the transformations between the field variables are given by
\begin{align*}
 &{\hat{q}}_1 = \frac{q_1\chi_1^2+3i\zeta_2\chi_1\chi_2+\zeta_2\chi_1R_2(q_1\chi_3-\chi_2q_2)}{\chi_2^2},\;\;
 \;\;\;\;
  {\hat{r}}_1 = \frac{(r_1\chi_2+r_2\chi_3)\chi_2-3i\zeta_2\chi_1\chi_2}{\chi_1^2},& \\
 & {\hat{r}}_2 =
\frac{(3iR_2\zeta_2^2-r_2)\chi_1\chi_2-\zeta_2\chi_2R_2(r_1\chi_2+r_2\chi_3)}{\zeta_2\chi_1^2},
\;\;\;
  {\hat{q}}_2 =\frac{(q_1\chi_3-q_2\chi_2)\zeta_2\chi_1}{\chi_2^2}.&
\end{align*}
\end{thm}
\noindent
\textbf{Proof:} Direct calculation.

Similar to the \textbf{Remark 1}, we have
\begin{rem}\label{rm:2}
An alternative formula for $\hat{d}$ is $\hat{d}_x={r}_2$. Thus, $\hat{d}=d$.
\end{rem}

Finally we may have a combined Darboux transformation in the
following manner: we take a particular solution
$\Phi_1\equiv(\varphi_1,\varphi_2,\varphi_3)^T$ of
\eqref{eq:zero1}-\eqref{eq:zero2} at $\zeta=\zeta_1$ and a
particular solution $(\chi_1,\chi_2,\chi_3)$ of
\eqref{con1}-\eqref{con2} at $\zeta=\zeta_2$. Then, with
$(\varphi_1,\varphi_2,\varphi_3)^T$ we may use \textbf{Theorem 1}
and have a Darboux transformation whose Darboux matrix is $T_1$. At
this stage, $(\chi_1,\chi_2,\chi_3)$ is converted into a new
solution $\Psi_1\equiv
(\hat{\chi}_1,\hat{\chi}_2,\hat{\chi}_3)=(\chi_1,\chi_2,\chi_3)T^{-1}_1|_{\zeta=\zeta_2}$
for the conjugate linear system. This solution, with the help of
\textbf{Theorem 2}, enables us to construct a Darboux matrix $T_2$
and take a Darboux transformation for the conjugate linear system,
which in turn induces a transformation for the original linear
system. Schematically it looks as
\[
\begin{CD}
{\Phi} @> T_1>{\rm seed:}\Phi_1> \hat{\Phi}
@>T_2^{-1}>\rm{seed:}\Psi_1> \Phi[1]
\end{CD}
\]
It is now easy to find the explicit formulae. Indeed, the three
components of $\Psi_1$ reads as
\begin{eqnarray*}
\hat{\chi}_1&=&\frac{\zeta_1\zeta_2\chi_1\varphi_1+\zeta_1^2\chi_2\varphi_2+\zeta_1^2\chi_3\varphi_3}{(\zeta_1^2-\zeta_2^2)\varphi_2},\\
\hat{\chi}_2&=&\frac{\zeta_1^2\chi_1\varphi_1+\zeta_1\zeta_2\chi_2\varphi_2+\zeta_1\zeta_2\chi_3\varphi_3}{(\zeta_1^2-\zeta_2^2)\varphi_1},\\
\hat{\chi}_3&=&\frac{\zeta_1(\zeta_1\chi_1\varphi_1+\zeta_2\chi_2\varphi_2+\zeta_2\chi_3\varphi_3)}{(\zeta_1^2-\zeta_2^2)\varphi_1}Q_2
                    +\frac{\chi_3}{\zeta_2}.
\end{eqnarray*}
Using this seed solution, we find that the functions appeared in
$T_2$ in the present case read as
\begin{eqnarray*}
  \hat{a} &=&
  -\frac{\varphi_2(\zeta_1\chi_1\varphi_1+\zeta_2\chi_2\varphi_2+\zeta_2\chi_3\varphi_3)}
  {\zeta_2\varphi_1(\zeta_2\chi_1\varphi_1+\zeta_1\chi_2\varphi_2+\zeta_1\chi_3\varphi_3)}, \\
  \hat{b} &=& \frac{\varphi_3}{\varphi_2} Q_2-
  \frac{\varphi_1(\zeta_1\zeta_2\chi_1\varphi_1+\zeta_1^2\chi_2\varphi_2+\zeta_2^2\chi_3\varphi_3)}
  {\zeta_1\zeta_2\varphi_2(\zeta_1\chi_1\varphi_1+\zeta_2\chi_2\varphi_2+\zeta_2\chi_3\varphi_3)}, \\
  \hat{c} &=&-Q_2+\frac{(\zeta_2^2-\zeta_1^2)\chi_3\varphi_1}
  {\zeta_1\zeta_2(\zeta_1\chi_1\varphi_1+\zeta_2\chi\varphi_2+\zeta_2\chi_3\varphi_3)},\\
  \hat{d}&=&-\frac{\varphi_3}{\varphi_2}.
\end{eqnarray*}
The Darboux matrix we are seeking, $T=T_2^{-1}T_1$, which after
removing an overall factor $\frac{\zeta_2^2}{(\zeta^2-\zeta_2^2)}$,
is
\begin{equation}\label{eq:T}
    T=\begin{pmatrix}
       a\zeta^2-1 & c_1\zeta & c_2\zeta \\
       c_3\zeta & b\zeta^2-1 & c\zeta^2 \\
       c_4\zeta & d\zeta^2 & e\zeta^2-1 \\
     \end{pmatrix},
\end{equation}
where
\begin{align}
&a=\frac{D_2}{\zeta_1\zeta_2D_1}, \;\;\;\;
b=\frac{D_3}{\zeta_1\zeta_2^2D_2},\;\;\;\;
e=\frac{D_4}{\zeta_1\zeta_2^2D_2},\;&\\
&c=\frac{(\zeta_2^2-\zeta_1^2)\chi_3\varphi_2}{\zeta_1\zeta_2^2D_2},\;\;\;\;\;
d=\frac{(\zeta_2^2-\zeta_1^2)\chi_2\varphi_3}{\zeta_1\zeta_2^2D_2},\;&\\
\label{eq18}
&c_1=\frac{(\zeta_2^2-\zeta_1^2)\chi_2\varphi_1}{\zeta_1\zeta_2D_1},\;\;\;
c_2=\frac{(\zeta_2^2-\zeta_1^2)\chi_3\varphi_1}{\zeta_1\zeta_2D_1},\;&\\
\label{eq19}
&c_3=\frac{(\zeta_2^2-\zeta_1^2)\chi_1\varphi_2}{\zeta_1\zeta_2D_2},\;\;\;
c_4=\frac{(\zeta_2^2-\zeta_1^2)\chi_1\varphi_3}{\zeta_1\zeta_2D_2},\;&
\end{align}
and
\begin{eqnarray*}
D_1&=&\zeta_1\chi_1\varphi_1+\zeta_2\chi_2\varphi_2+\zeta_2\chi_3\varphi_3,\\
D_2&=&\zeta_2\chi_1\varphi_1+\zeta_1\chi_2\varphi_2+\zeta_1\chi_3\varphi_3,\\
D_3&=&\zeta_1\zeta_2\chi_1\varphi_1+\zeta_2^2\chi_2\varphi_2+\zeta_1^2\chi_3\varphi_3,\\
D_4&=&\zeta_1\zeta_2\chi_1\varphi_1+\zeta_1^2\chi_2\varphi_2+\zeta_2^2\chi_3\varphi_3.
\end{eqnarray*}

The transformations between field variables can be reformed neatly
%
\begin{eqnarray} \label{eq:r1}
{q}_1[1] = q_1-c_{1,x}, &&
{q}_2[1] = q_2-c_{2,x}, \\
\label{eq:r2} {{r}}_1[1] = r_1 -c_{3,x}, && {{r}}_2[1]=
r_2-c_{4,x},
\end{eqnarray}
and $c_i$'s are given by \eqref{eq18}-\eqref{eq19}.

\section{Reduction}
In last section, we constructed a combined or two-fold Darboux
transformation for our linear system
\eqref{eq:zero1}-\eqref{eq:zero2}. The relevant Darboux matrix and
field variable transformations are given by \eqref{eq:T} and
\eqref{eq:r1}-\eqref{eq:r2} respectively. What we are interested in
is to present a Darboux transformation for the two component DNLS
equation and thus we have to do reduction.  Next we will show that
our Darboux transformation can be reduced easily to the interested
case.

The constraints between field variables are
\[
r_1=-q_1^*,\;\; r_2=-q_2^*
\]
which should be kept invariant under Darboux transformation.
Now we notice that, for the solution ~$(\varphi_1,\varphi_2,\varphi_3)^T$ of the linear system  \eqref{eq:zero1}-\eqref{eq:zero1} at $\zeta=\zeta_1$,
~$(\varphi_1^*,\varphi_2^*,\varphi_3^*)$ is the solution of conjugate linear system equation \eqref{con1}-\eqref{con2} at~$\zeta=\zeta_1^*$. Therefore, we use it as our seed for the second step Darboux transformation. Namely,
\[
 \Psi_1=(\varphi_1^*,\varphi_2^*,\varphi_3^*),\;\;\zeta_2=\zeta_1^*.
\]
With these considerations, it is easy to verify that
\[
c_1^*=-c_3,\;\; c_2^*=c_4
\]
therefore
\[
r_1[1]=-q_1[1]^*,\;\; r_2[1]=-q_2[1]^*
\]
The final transformation is neatly written as
\begin{eqnarray}\label{eq:d}
  {q}_1[1] &=& q_1-\frac{\zeta_1^{*2}-\zeta_1^2}{{|\zeta_1^2|}}\left(\frac{\varphi_1\varphi_2^*}{|\varphi_1^2|\zeta_1+\zeta_1^*(|\varphi_2^2|+|\varphi_3^2|)}\right)_x ,\\
\label{eq:d1}
{q}_2[1] &=&
q_2-\frac{\zeta_1^{*2}-\zeta_1^2}{{|\zeta_1^2|}}\left(\frac{\varphi_1\varphi_3^*}{|\varphi_1^2|\zeta_1+\zeta_1^*(|\varphi_2^2|+|\varphi_3^2|)}\right)_x.
\end{eqnarray}

If we start with the vacuum solution $q_1=q_2=0$, then the linear system \eqref{eq:zero1}-\eqref{eq:zero2} has a solution
\[
\varphi_1=e^{-2i\zeta_1^2x-9i\zeta_1^4t},\quad \varphi_2=e^{i\zeta_1^2x}=\varphi_3
\]
which leads to
\begin{eqnarray*}
 {{q}}_1[1]={{q}}_2[1]=
  \frac{6\zeta_1^2\mathrm{Im}(\zeta_1^2)e^{-iR}\left[\sinh(I)(\zeta_1^{*2}-2|\zeta_1^2|)+\cosh(I)(\zeta_1^{*2}+2|\zeta_1^2|)\right]}
  {\zeta_1^*\left[\sinh(I)(\zeta_1^{2}+2|\zeta_1^2|)+\cosh(I)(\zeta_1^{2}-2|\zeta_1^2|)\right]^2} \\
\end{eqnarray*}
where $R=3\mathrm{Re}(\zeta_1^2)x+9\mathrm{Re}(\zeta_1^4)t$,
$I=3\mathrm{Im}(\zeta_1^2)x+9\mathrm{Im}(\zeta_1^4)t$. It is nothing but  a solution of the DNLS equation. To find more interesting ones we need to iterate our Darboux transformation and we will do so in next section.

\section{Iterations: N-fold Darboux matrix}
The appealing feature of a Darboux transformation is that it often leads to determinant representation for $N-$solitons. To this aim,
one has to do iteration. In this section, we consider the iteration problem for our Darboux transformation.

First, let us rewrite our Darboux matrix  $T$ given by (\ref{eq:T}) with the reductions in mind. Introduce a new matrix
\[
N(\zeta)=\mathrm{diag}(\frac{\zeta\varphi_1}{\zeta_1D},\frac{\zeta\varphi_2}{\zeta_1D^*},\frac{\zeta\varphi_3}{\zeta_1D^*})
\]
where
$D=D_1|_{\chi_j=\varphi_j^*,\zeta_2=\zeta_1^*}$.
 Then, the Darboux matrix $T$ takes the following form
\begin{eqnarray*}
 T=\frac{\zeta^2-\zeta_1^{*2}}{\zeta_1^{*2}} +\frac{\zeta_1^{*2}-\zeta_1^2}{\zeta_1^{*2}}N(\zeta)
                                                                \begin{pmatrix}
                                                                  \zeta\varphi_1^* & \zeta_1^*\varphi_2^* & \zeta_1^*\varphi_3^* \\
                                                                  \zeta_1^*\varphi_1^* & \zeta\varphi_2^* & \zeta\varphi_3^* \\
                                                                  \zeta_1^*\varphi_1^* & \zeta\varphi_2^* & \zeta\varphi_3^* \\
                                                                \end{pmatrix}.
\end{eqnarray*}
Now, on the one hand we have already known
\begin{equation}\label{eq:11}
T|_{\zeta=\zeta_1}\begin{pmatrix}
                                     \varphi_1\\ \varphi_2 \\ \varphi_3\\
                                    \end{pmatrix}=0.
\end{equation}
i.e. our seed $\Phi_1=(\varphi_1,\varphi_2,\varphi_3)^T$ lies in the kernal of the matrix $T|_{\zeta=\zeta_1}$. On the other hand, let us suppose
\begin{equation}\label{eq:12}
T|_{\zeta=\zeta_1^*}\begin{pmatrix}
                                     \psi_1\\ \psi_2 \\ \psi_3\\
                                    \end{pmatrix}
=\frac{\zeta_1^{*2}-\zeta_1^2}{\zeta_1^{*}}N(\zeta_1^*)
                                                                \begin{pmatrix}
                                                                  \varphi_1^*\psi_1+\varphi_2^*\psi_2+\varphi_1^*\psi_3\\
                                                                  \varphi_1^*\psi_1+\varphi_2^*\psi_2+\varphi_1^*\psi_3\\
                                                                  \varphi_1^*\psi_1+\varphi_2^*\psi_2+\varphi_1^*\psi_3\\
                                                                \end{pmatrix}=0
\end{equation}
for certain vector function  $\Psi_1=(\psi_1,\psi_2,\psi_3)^T$, then for $\zeta_1\neq \zeta_1^*$ one has to impose $\varphi_1^*\psi_1+\varphi_2^*\psi_2+\varphi_1^*\psi_3=0$, or
\[
\Phi_{1}^{\dag}\Psi_1=0,
\]
obviously
\[
\Psi_{1}^{1}=(-\varphi_2^*,\varphi_1^*,0)^{T},
\Psi_{1}^{2}=(-\varphi_3^*,0,\varphi_1^*)^{T}
\]
meet the requirment
\begin{equation}\label{eq-seed}
T|_{\zeta=\zeta_{1}^*}\Psi_{1}^{k}=0.
\end{equation}
We observe that the conditions \eqref{eq:11} and \eqref{eq-seed} can in turn be   used to determine the nine quantities appeared in $T$ \emph{uniquely}.

Now we are ready to do iterations.  Assume that we are given $N$
distinct complex numbers $\zeta_1,\zeta_2,...,\zeta_N$
 such that $\zeta_k^{*2}\neq\zeta_k^2$ $(k=1, 2, ..., N)$. We further assume that the
 vector
\[
\Phi_k =(\varphi_1^{(k)},\varphi_2^{(k)},\varphi_3^{(k)})^T
\]
is a solution of linear equation at $\zeta=\zeta_k$, i.e.
\begin{equation*}
    [\partial_x-U(\zeta=\zeta_k)](\Phi_k)=0,\qquad
    [\partial_t-V(\zeta=\zeta_k)](\Phi_k)=0,
\end{equation*}
and
\[
\Psi_k^1=( -\varphi_2^{(k)*},\varphi_1^{(k)*},0)^T,\quad
\Psi_k^2=(-\varphi_3^{(k)*},0,\varphi_1^{(k)*})^T,
\]
which  satisfy the orthogonal conditions $\Phi_k^{\dag}\Psi_k^l=0$.

With these seed solutions, we define
\begin{eqnarray*}
T_k=\frac{\zeta^2-\zeta_k^{*2}}{\zeta_k^{*2}}+\frac{\zeta_k^{*2}-\zeta_k^2}{\zeta_k^{*2}}N_k(\zeta)
\begin{pmatrix}
\zeta\varphi_1^{(k)}[k-1]^* & \zeta_k^*\varphi_2^{(k)}[k-1]^* & \zeta_k^*\varphi_3^{(k)}[k-1]^* \\
\zeta_i^*\varphi_1^{(k)}[k-1]^* & \zeta\varphi_2^{(k)}[k-1]^* & \zeta\varphi_3^{(k)}[k-1]^* \\
\zeta_i^*\varphi_1^{(k)}[k-1]^* & \zeta\varphi_2^{(k)}[k-1]^* & \zeta\varphi_3^{(k)}[k-1]^* \\
\end{pmatrix},
\end{eqnarray*}
where
\begin{eqnarray*}
D[k]&=&\zeta_k\left|\varphi_1^{(k)}[k-1]\right|^2+\zeta_k^*(\left|\varphi_2^{(k)}[k-1]\right|^2+\left|\varphi_3^{(k)}[k-1]\right|^2),\\
N_k(\zeta)&=&\textrm{diag}\left(\frac{\zeta\varphi_1^{(k)}[k-1]^*}{\zeta_iD[k]},
\frac{\zeta\varphi_2^{(k)}[k-1]^*}{\zeta_iD[k]^*},\frac{\zeta\varphi_3^{(k)}[k-1]^*}{\zeta_iD[k]^*}\right),
\end{eqnarray*}
and our notation is the following
\begin{eqnarray*}
\Phi_j[k]=\begin{pmatrix}
            \varphi_1^{(j)}[k] \\
            \varphi_2^{(j)}[k] \\
            \varphi_3^{(j)}[k] \\
          \end{pmatrix}
=T_kT_{k-1}...T_1|_{\zeta=\zeta_j}\begin{pmatrix}\varphi_1^{(j)}\\\varphi_2^{(j)}\\\varphi_3^{(j)}\end{pmatrix},
\end{eqnarray*}
and  $\Phi_j[0]=\Phi_j$.

 The $N$-times iterated Darboux
matrix is given by
\begin{eqnarray*}
  T &=& T_NT_{N-1}\cdots T_1.
\end{eqnarray*}

It is easy to see that, similar to the equation (\ref{eq:11}), the
following relations hold
\begin{eqnarray*}
  T_kT_{k-1}\cdots T_1|_{\zeta=\zeta_k}\Phi_k &=& T_k|_{\zeta=\zeta_k}\Phi_k[k-1]=0\qquad
(k=1, 2, ..., N).
\end{eqnarray*}

Furthermore, we recursively define
\[
\Psi_k^{l}[0]=\Psi_k^l,\quad \Psi_k^l[j-1]=T_{j-1}T_{j-2}\cdots
T_1|_{\zeta=\zeta_{k}^*}\Psi_k^l,
\]
then we have
\begin{prop}\label{lem:1}
$\Phi_k^{\dag}[k-1]\Psi_k^l[k-1]=0$.
\end{prop}
\noindent \textbf{Proof:} We know $\Phi_k^{\dag}[0]\Psi_k^{l}[0]=0$.
Let us suppose $\Phi_k^{\dag}[m]\Psi_k^{l}[m]=0$ ($0\leq m<k-1$).
Then thanks to $\Phi_k[m+1]=T_{m+1}|_{\zeta=\zeta_k}\Phi_k[m]$ and
$\Psi_k^{l}[m+1]=T_{m+1}|_{\zeta=\zeta_k^*}\Psi_k^{l}[m]$, we have
\[
\Psi_k^{\dag}[m+1]\Phi_k^l[m+1]=\Psi_k^{\dag}[m]T_{m+1}^{\dag}|_{\zeta=\zeta_{k}}T_{m+1}|_{\zeta=\zeta_{k}^*}\Phi_k^l[m]=0,
\]
because
 of $T_{m+1}^{\dag}|_{\zeta=\zeta_{k}}T_{m+1}|_{\zeta=\zeta_{k}^*}
=\frac{1}{|\zeta_{m+1}^4|}(\zeta_{k}^{*2}-\zeta_{m+1}^{*2})(\zeta_{k}^{*2}-\zeta_{m+1}^2)$.
Therefore, the lemma follows from the mathematical induction.

\bigskip

Based on \textbf{Proposition 1}, we obtain
\begin{eqnarray*}
  T_kT_{k-1}...T_1|_{\zeta=\zeta_k^*}\Psi_k^l &=&
  T_k|_{\zeta=\zeta_k^*}\Psi_k^l[k-1]=0,\qquad(l=1,2).
\end{eqnarray*}
Therefore, we have
\begin{equation}
\label{eq:13}  T|_{\zeta=\zeta_k}\Phi_k = 0,\quad  T|_{\zeta=\zeta_k^*}\Psi_k^1 = 0,\quad  T|_{\zeta=\zeta_k^*}\Psi_k^2 = 0,
\end{equation}
for $k=1, 2, ..., N$.
We also notice that our iterated Darboux matrix $T$ is taking of the form
\begin{displaymath}
   T=\sum_{k=0}^{2N} \zeta^{k}T_{k}
     = \sum\limits_{n=1}^{N}\begin{pmatrix}
        a_{2n}\zeta^{2n}&   c_1^{(2n-1)}\zeta^{2n-1} &   c_2^{(2n-1)}\zeta^{2n-1} \\[6pt]
        c_3^{(2n-1)}\zeta^{2n-1} &  b_{2n}\zeta^{2n} &  c_{2n}\zeta^{2n} \\[6pt]
        c_4^{(2n-1)}\zeta^{2n-1} &  d_{2n}\zeta^{2n} &  e_{2n}\zeta^{2n}\\
     \end{pmatrix}+ (-1)^N.
\end{displaymath}
Above coefficients can be determinated in $T$ by solving the linear
algebraic systems (\ref{eq:13}).   The solution formulae are
obtained from
\begin{displaymath}\label{N-sol}
{q}_1[N] = q_1+ \left(\frac{H_2}{H_1}\right)_x ,\qquad {q}_2[N]  =
q_2+
\left(\frac{H_3}{H_1}\right)_x,
\end{displaymath}
where
\begin{eqnarray*}
  H_1 &=& \begin{vmatrix}
                  \zeta_1^{2N}\varphi_1^{(1)} &  \zeta_1^{2N-1}\varphi_2^{(1)} &  \zeta_1^{2N-1}\varphi_3^{(1)} &...& \zeta_1^2\varphi_1^{(1)} &  \zeta_1\varphi_2^{(1)} &  \zeta_1\varphi_3^{(1)} \\[5pt]
                   -\zeta_1^{2N*}\varphi_2^{(1)*} &  \zeta_1^{2N-1*}\varphi_1^{(1)*} &  0 & ... &  -\zeta_1^{2*}\varphi_2^{(1)*} & \zeta_1^{*}\varphi_1^{(1)*} &  0\\[5pt]
                   -\zeta_1^{2N*}\varphi_3^{(1)*} &  0 &  \zeta_1^{2N-1*}\varphi_1^{(1)*} & ... &  -\zeta_1^{2*}\varphi_3^{(1)*} &   0 & \zeta_1^{*}\varphi_1^{(1)*} \\[5pt]
                 ... & ... & ... &... & ... & ... &... \\[5pt]
                  \zeta_N^{2N}\varphi_1^{(N)} &  \zeta_N^{2N-1}\varphi_2^{(N)} &  \zeta_N^{2N-1}\varphi_3^{(N)} &  ...&  \zeta_N^2\varphi_1^{(N)} &  \zeta_N\varphi_2^{(N)} &  \zeta_N\varphi_3^{(N)} \\[5pt]
                  -\zeta_N^{2N*}\varphi_2^{(N)*} &  \zeta_n^{2N-1*}\varphi_1^{(N)*} &  0 & ... &  -\zeta_N^{2*}\varphi_2^{(N)*} & \zeta_N^{*}\varphi_1^{(N)*} &  0\\[5pt]
                  -\zeta_N^{2N*}\varphi_3^{(N)*} &  0 &  \zeta_n^{2N-1*}\varphi_1^{(N)*} & ... &  -\zeta_N^{2*}\varphi_3^{(N)*} &  0 & \zeta_N^{*}\varphi_1^{(N)*} \\[5pt]
               \end{vmatrix},
\end{eqnarray*}
and
~$H_2$ and ~$H_3$ are  $H_1$ with the $3N-1\textrm{th}$ column and the $3N$th
column replaced by $L_1$ respectively. Where
\begin{displaymath}
L_1=\begin{pmatrix}
    -\varphi_1^{(1)}, &
   \varphi_2^{(1)*}, &
   \varphi_3^{(1)*}, &
   ..., &
    -\varphi_1^{(N)}, &
    \varphi_2^{(N)*}, &
    \varphi_3^{(N)*}, &
 \end{pmatrix}^T,
\end{displaymath}

To demonstrate the usefulness our solution formulae, we calculate solutions for the two component DNLS equation.
Selecting
\[
\zeta_1=1+\frac{1}{3}i, \zeta_2=1+\frac{2}{3}i,
\Phi_1=(e^{-2i\zeta_1^2x-9i\zeta_1^4t},0,e^{i\zeta_1^2x})^T,
\Phi_2=(e^{-2i\zeta_2^2x-9i\zeta_2^4t},e^{i\zeta_2^2x},e^{i\zeta_2^2x})^T,
\]
and substituting them into \eqref{N-sol} we could have the
solutions. Figure 1 and Figure 2 show these solutions by plotting
$|q_1^2|$ and $|q_2^2|$. It is pointed out that while the second
figure exhibits standard two-soliton scattering, the first one
demonstrates a fission process.
\begin{figure}
        \centering
        \begin{minipage}[c]{0.4\linewidth}
          \centering
       \includegraphics[width=4.0in]{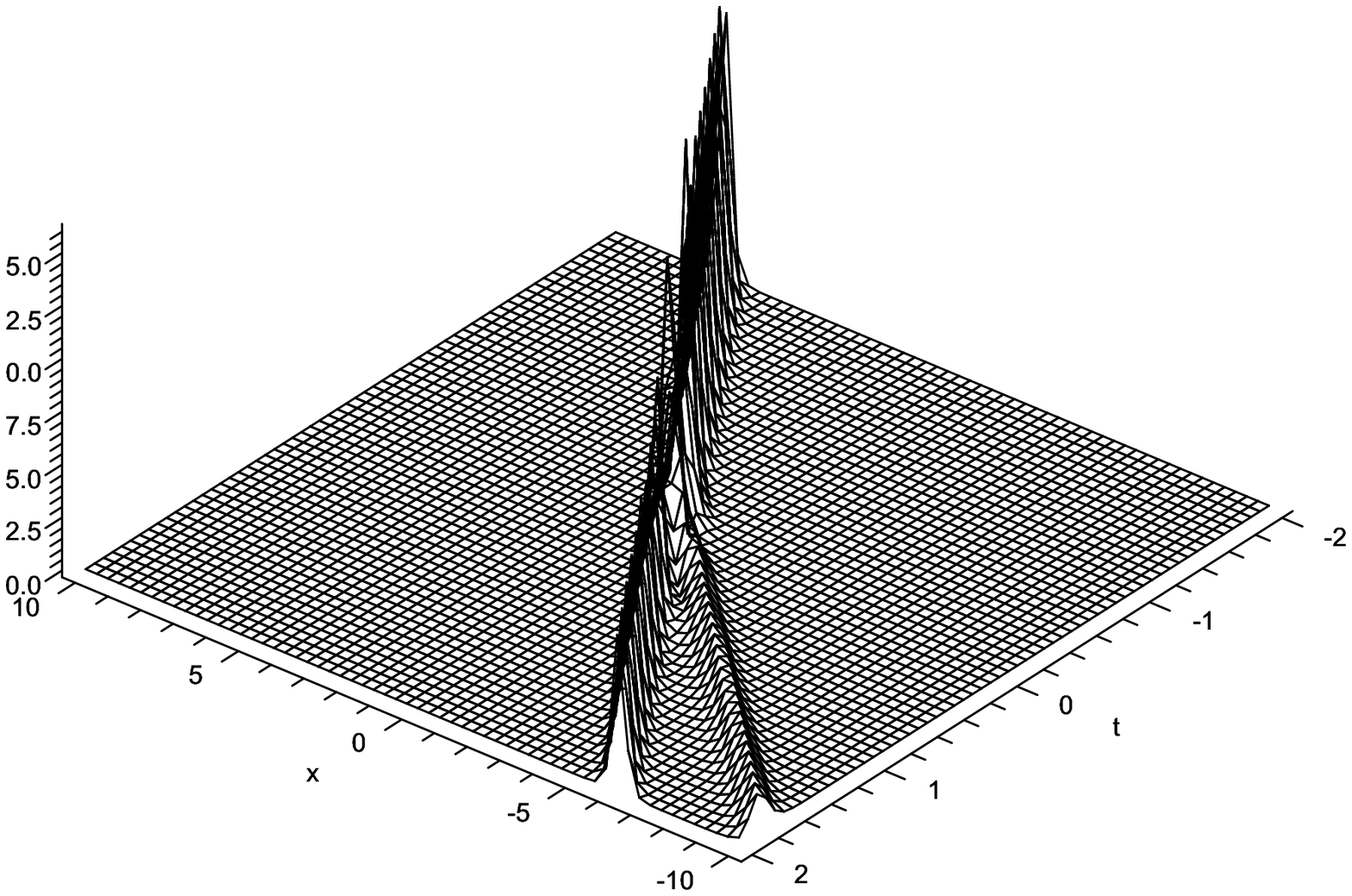}
          \caption{$|q_1^2|$}
        \end{minipage}%
\hfill \centering
        \begin{minipage}[c]{0.4\linewidth}
         \centering
       \includegraphics[width=4.0in]{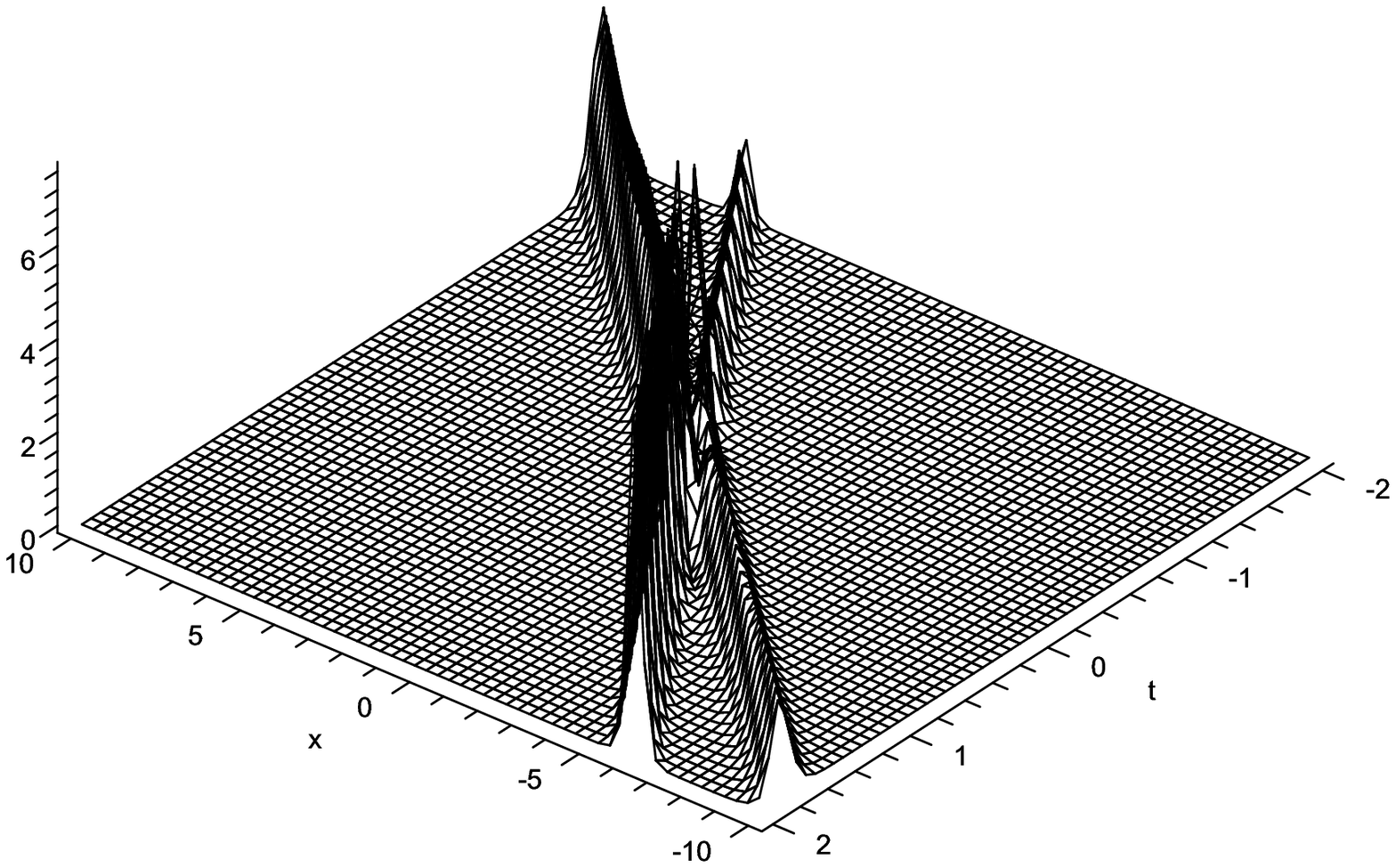}
          \caption{$|q_2^2|$}
        \end{minipage}%
      \end{figure}

\section{Conclusion}
Above we found a Darboux transformation for the two component DNLS equation and obtained a closed formula for its solutions.
We remark that our Darboux transformation can be easily generalized to multi component case. In fact, the Darboux matrix
in this case is
\begin{eqnarray*}
 T=\frac{\zeta^2-\zeta_1^{*2}}{\zeta_1^{*2}}  +\frac{\zeta_1^{*2}-\zeta_1^2}{\zeta_1^{*2}}N(\zeta)\begin{pmatrix}
                                                               \zeta\varphi_1^* & \zeta_1^*\varphi_2^* & \zeta_1^*\varphi_3^*&...&\zeta_1^*\varphi_n^* \\
                                                                 \zeta_1^*\varphi_1^* & \zeta\varphi_2^* & \zeta\varphi_3^* &...&\zeta\varphi_n^*\\
                                                                 \zeta_1^*\varphi_1^* & \zeta\varphi_2^* & \zeta\varphi_3^* &...&\zeta\varphi_n^*\\
                                                                ... & ... & ... & ... & ... \\
                                                                 \zeta_1^*\varphi_1^* & \zeta\varphi_2^* & \zeta\varphi_3^* &...&\zeta\varphi_n^*\\
                                                                \end{pmatrix},
\end{eqnarray*}
where
\begin{eqnarray*}
N(\zeta)=\mathrm{diag}(\frac{\zeta\varphi_1}{\zeta_1D},\frac{\zeta\varphi_2}{\zeta_1D^*},\cdots,\frac{\zeta\varphi_n}{\zeta_1D^*}),
\end{eqnarray*}
with
\[
D=\zeta_1|\varphi_1^2|+\zeta_1^*|\varphi_2^2|+...+\zeta_1^*|\varphi_n^2|.
\]
and solution formulae may be derived.

\textbf{Acknowledgment} The work is supported by the National Natural Science Foundation of China (grant numbers: 10671206, 10731080, 10971222).

\end{document}